\documentclass[submitting]{nst}

\usepackage{subfigure,dcolumn}
\usepackage[T2A,T1]{fontenc}
\usepackage[russian,english]{babel}

\usepackage{listings}
\begin{document}

\title{Development of a low-background neutron detector array}

\author{Y. T. Li}
\affiliation{CAS Key Laboratory of High Precision Nuclear Spectroscopy, Institute of Modern Physics, Chinese Academy of Sciences, Lanzhou 730000, China}
\affiliation{School of Nuclear Science and Technology, University of Chinese Academy of Sciences, Beijing 100049, China}
\author{W. P. Lin}
\email[Corresponding author, ]{linwp1204@scu.edu.cn}
\affiliation{Key Laboratory of Radiation Physics and Technology of the Ministry of Education, Institute of Nuclear Science and Technology, Sichuan University, Chengdu 610064, China}
\author{B. Gao}
\email[Corresponding author, ]{gaobsh@impcas.ac.cn}
\affiliation{CAS Key Laboratory of High Precision Nuclear Spectroscopy, Institute of Modern Physics, Chinese Academy of Sciences, Lanzhou 730000, China}
\affiliation{School of Nuclear Science and Technology, University of Chinese Academy of Sciences, Beijing 100049, China}
\affiliation{Joint Department for Nuclear Physics, Lanzhou University and Institute of Modern Physics, Chinese Academy of Sciences, Lanzhou 730000, China}
\author{H. Chen}
\affiliation{CAS Key Laboratory of High Precision Nuclear Spectroscopy, Institute of Modern Physics, Chinese Academy of Sciences, Lanzhou 730000, China}
\author{H. Huang}
\affiliation{Joint Department for Nuclear Physics, Lanzhou University and Institute of Modern Physics, Chinese Academy of Sciences, Lanzhou 730000, China}
\author{Y. Huang}
\affiliation{Key Laboratory of Radiation Physics and Technology of the Ministry of Education, Institute of Nuclear Science and Technology, Sichuan University, Chengdu 610064, China}
\author{T. Y. Jiao}
\affiliation{CAS Key Laboratory of High Precision Nuclear Spectroscopy, Institute of Modern Physics, Chinese Academy of Sciences, Lanzhou 730000, China}
\affiliation{School of Nuclear Science and Technology, University of Chinese Academy of Sciences, Beijing 100049, China}
\author{K. A. Li}
\affiliation{CAS Key Laboratory of High Precision Nuclear Spectroscopy, Institute of Modern Physics, Chinese Academy of Sciences, Lanzhou 730000, China}
\affiliation{School of Nuclear Science and Technology, University of Chinese Academy of Sciences, Beijing 100049, China}
\author{X. D. Tang}
\affiliation{CAS Key Laboratory of High Precision Nuclear Spectroscopy, Institute of Modern Physics, Chinese Academy of Sciences, Lanzhou 730000, China}
\affiliation{School of Nuclear Science and Technology, University of Chinese Academy of Sciences, Beijing 100049, China}
\affiliation{Joint Department for Nuclear Physics, Lanzhou University and Institute of Modern Physics, Chinese Academy of Sciences, Lanzhou 730000, China}
\author{X. Y. Wang}
\affiliation{CAS Key Laboratory of High Precision Nuclear Spectroscopy, Institute of Modern Physics, Chinese Academy of Sciences, Lanzhou 730000, China}
\affiliation{School of Nuclear Science and Technology, University of Chinese Academy of Sciences, Beijing 100049, China}
\author{X. Fang}
\affiliation{Sino-French Institute of Nuclear Engineering and Technology,
Sun Yat-sen University, Zhuhai, Guangdong 519082, People’s Republic of China}
\author{H. X. Huang}
\affiliation{China Institute of Atomic Energy, Beijing 102413, People’s Republic of China}
\author{J. Ren}
\affiliation{China Institute of Atomic Energy, Beijing 102413, People’s Republic of China}
\author{L. H. Ru}
\affiliation{CAS Key Laboratory of High Precision Nuclear Spectroscopy, Institute of Modern Physics, Chinese Academy of Sciences, Lanzhou 730000, China}
\affiliation{School of Nuclear Science and Technology, University of Chinese Academy of Sciences, Beijing 100049, China}
\author{X. C. Ruan}
\affiliation{China Institute of Atomic Energy, Beijing 102413, People’s Republic of China}
\author{N. T. Zhang}
\affiliation{CAS Key Laboratory of High Precision Nuclear Spectroscopy, Institute of Modern Physics, Chinese Academy of Sciences, Lanzhou 730000, China}
\affiliation{School of Nuclear Science and Technology, University of Chinese Academy of Sciences, Beijing 100049, China}
\author{Z. C. Zhang}
\affiliation{CAS Key Laboratory of High Precision Nuclear Spectroscopy, Institute of Modern Physics, Chinese Academy of Sciences, Lanzhou 730000, China}
\affiliation{School of Nuclear Science and Technology, University of Chinese Academy of Sciences, Beijing 100049, China}

\begin{abstract}
 A low-background neutron detector array was developed to measure
the cross section of the $^{13}$C($\alpha$,n)$^{16}$O reaction, which is the
neutron source for the $s$-process in AGB stars, in the
Gamow window ($E_{c.m.}$ = 190 $\pm$ 40 keV) at the China Jinping
Underground Laboratory (CJPL).  The detector array consists of 24
$^{3}$He proportional counters embedded in a polyethylene cube.  Due to
the deep underground location and a borated polyethylene shield around
the detector array, a low background of 4.5(2)/hour was achieved.  The
$^{51}$V(p, n)$^{51}$Cr reaction was used to determine the neutron
detection efficiency of the array for neutrons with energy $E_n$ $<$ 1 MeV. Geant4 simulations, which were shown
to well reproduce experimental results, were used to extrapolate the
detection efficiency to higher energies for neutrons emitted in the $^{13}$C($\alpha$,n)
$^{16}$O reaction.
The theoretical angular distributions of the $^{13}$C($\alpha$,n)$^{16}$O
reaction were shown to be important in estimating the uncertainties of the detection
efficiency.
\end{abstract}

\keywords{Underground laboratory , Neutron detector , Low background , $^{3}$He Counter}

\maketitle

\section{Introduction}
\label{introduction}
The $^{13}$C($\alpha$,n)$^{16}$O reaction is the dominant neutron source for
the slow neutron capture process (s-process), which synthesized roughly half
of the elements heavier than iron in the Universe \cite{bib:1}.
 The main site of the s-process is the ``He intershell'' of asymptotic giant
 branch (AGB) stars where the temperature is 0.1 GK.
This corresponds to a Gamow window of 190 $\pm$ 40 keV for the $^{13}
$C($\alpha$,n)$^{16}$O reaction, which is far below the  Coulomb barrier.
Theoretical calculations predicted the cross section as low as
$\sim$ $10^{-14}$ barn at 190 keV \cite{bib:2,bib:3}.
Meanwhile, the $^{13}$C($\alpha$,n)$^{16}$O reaction is also the main neutron source for the intermediate neutron capture process (i-process) with a typical temperature 0.2 GK \cite{bib:4,bib:5}. A reliable direct measurement is still missing within the corresponding energy range from 0.2 to 0.54 MeV. 
To push direct measurements of the cross section towards the Gamow
window, high-intensity beams and low-background neutron detector array are
indispensable.


Present measurements performed at ground laboratories are limited by the
cosmic-ray-induced background.
With a typical background counting rate of a few hundred per hour (e.g.
\cite{bib:6,bib:7}), those measurements only reached
a lower limit of $E_\alpha$
$\sim$ 400 keV with a cross section of $\sim$ $10^{-10}$ barn.
The China JinPing underground Laboratory (CJPL) \cite{bib:8} with about 2400 m rock
overburden (6700 m water equivalent) shields against most of the cosmic
rays. The ultra-low background provides new opportunities for neutrino
physics \cite{bib:add1}, search of dark matter \cite{bib:add2, bib:add3} and direct cross
section measurements of key reactions in nuclear astrophysics \cite{bib:9,bib:10}.
The Jinping Underground Nuclear Astrophysics experimental facility (JUNA)
\cite{bib:9,bib:10} has
been initiated to measure important stellar nuclear reaction rates, taking
advantage of the low-background environments in CJPL.
The present JUNA project includes $^{12}$C($\alpha$,$\gamma$)$^{16}$O,
$^{13}$C($\alpha$,n)$^{16}$O, $^{19}$F(p,$\gamma$)$^{20}$Ne and $^{25}$Mg(p,
$\gamma$)$^{26}$Al reactions \cite{bib:9,bib:10}.
A high-intensity accelerator capable of delivering proton and $\alpha$ beams
up to 10 mA in the energy range of 0.05 - 0.4 MeV has been built for
JUNA \cite{bib:9}.
This made it possible to measure the $^{13}$C($\alpha$,n)$^{16}$O cross section at
the region of the s-process Gamow window if combined with a high-efficiency and low-background
neutron detection system.

This paper describes the development and characterization of a
low-background neutron detector array  for the cross section measurement of the
$^{13}$C($\alpha$, n)$^{16}$O reaction.

\section{Design of the detector array}
\begin{figure}[hbt]
	\centering
	\includegraphics[width=0.4\textwidth]{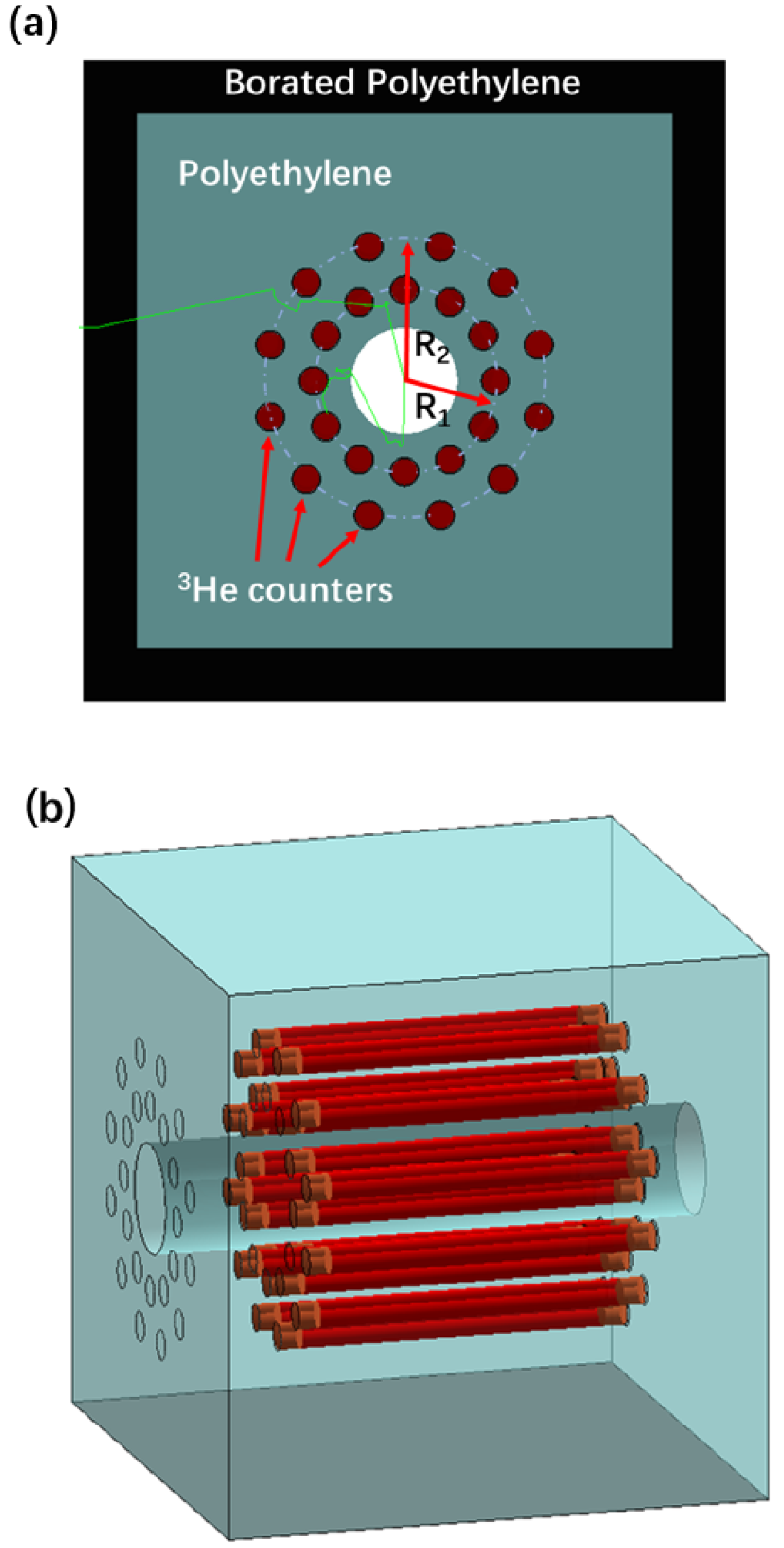}
	\caption{(Color online) Schematic view of the detector array. In panel (a), the borated
	polyethylene shield (black brick), positions of the $^{3}$He counters (red tubes) and the central
	bore hole are indicated. Traces of two neutron events are also shown as green curves. In
	panel (b), a three-dimensional view of the detector array is presented.}
	\label{fig:fig_detector_sketch}
\end{figure}

The detector array consists of 24 $^{3}$He-filled proportional counters
(manufactured by  GE-Reuter Stokes) embedded in a
 50 $\times$ 50 $\times$ 50 cm$^3$  polyethylene cube. Each proportional counter is filled with $^{3}$He and argon gases, with pressures of 4 bar and 2.4 bar, respectively.
The active length of each counter is 30 cm with a diameter of 2.54 cm.
The sketch of the detector array is shown in Fig. \ref{fig:fig_detector_sketch}.
The 24 proportional counters are evenly distributed in two concentric rings
with radius of $R_1$ = 8.5 cm and $R_2$ = 13 cm, respectively.
The radius of the two rings were optimized using Geant4 simulations,
following the same procedure as that in \cite{bib:11}, to get the maximum detection
efficiency for 2.0-MeV isotropic neutrons.
A central bore hole with a diameter of 10 cm was created to accommodate the
beam line and the target.
Neutrons from the $^{13}$C($\alpha$, n)$^{16}$O reaction with energies of 2 - 3
MeV are first moderated by the polyethylene cube and then captured by the
proportional counters.
To shield against environmental neutrons, the polyethylene cube is wrapped
with a 5 cm-thick 7\% borated polyethylene layer.

\begin{figure}[hbt]
\includegraphics[width=0.4\textwidth]{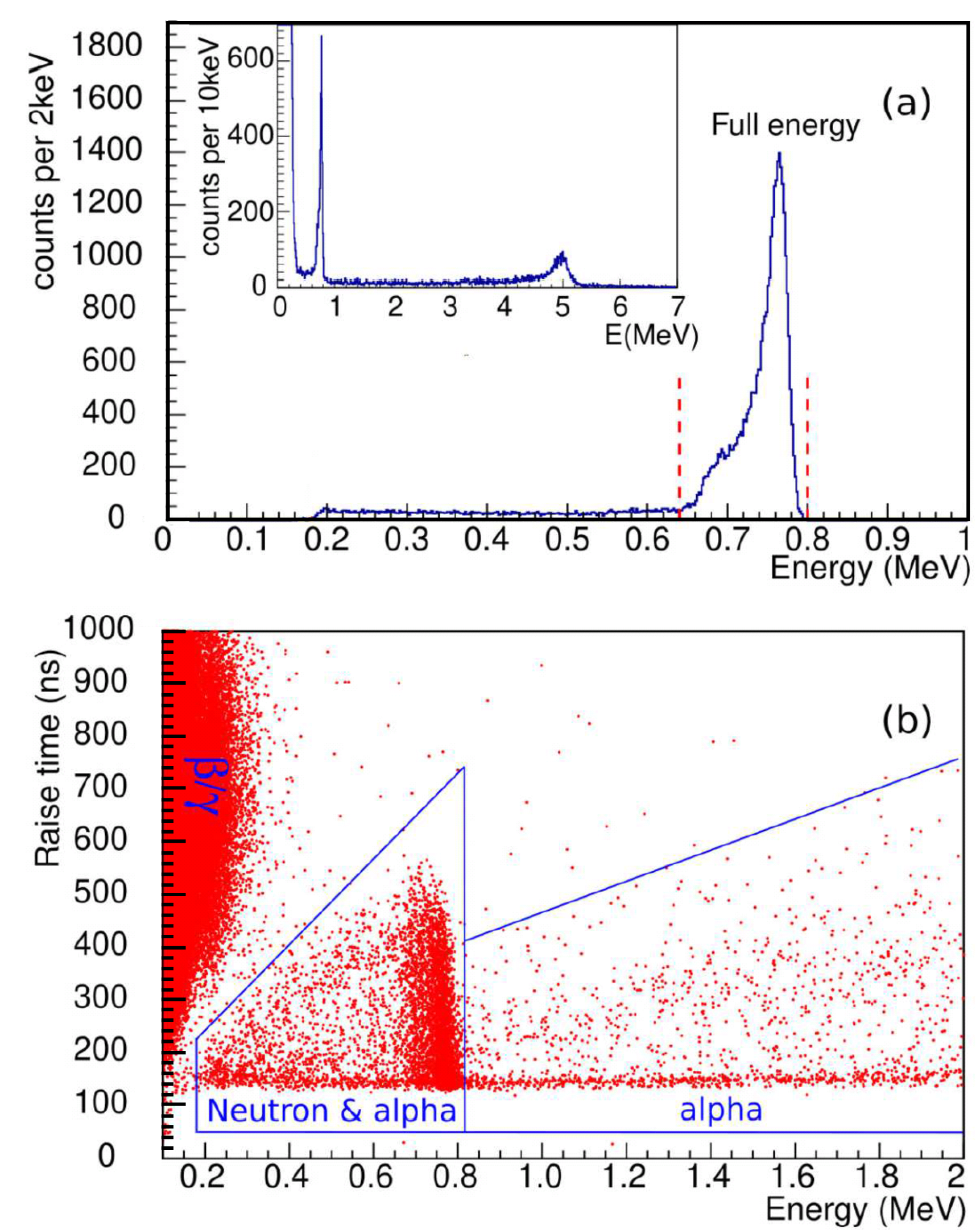}
	\caption{(Color online)
	(a) Typical energy spectrum measured by the detector. A cut as
	shown in panel (b) was applied to select the neutrons (see text). The dashed lines
	indicate the regions of interest between 0.64 - 0.8 MeV. The inset
	shows an energy spectrum in wider energy range without cuts.
	(b) Raise time versus energy two-dimensional
	plot measured by the detector array. The $\beta$/$\gamma$-, neutron \& alpha- and
	alpha-regions are indicated in the plot.
\label{fig_energy_spec}}
\end{figure}

The signal of each detector is processed by a charge-sensitive pre-amplifier (model
CAEN A1422). The output waveforms are digitized and recorded by a XIA Pixie-16 100 MHz card.
Fig. \ref{fig_energy_spec} (a) shows a typical energy spectrum measured
by the detector array using a $^{252}$Cf source.
The ``Neutron \& alpha'' cut as shown in Fig. \ref{fig_energy_spec} (b) was used in the plot to
select the neutron events (see the following sections for details).
The peak at 0.76 MeV corresponds to full-energy events where the triton and proton
produced from the $^{3}$He + $n$ $\rightarrow$ $p$ + $t$ reaction deposit
all their kinetic energies in
the sensitive region of the counter.
If the proton or triton hits the wall of the detector before depositing all their
kinetic energies in the sensitive region, the detector
records only part of the reaction $Q$ value.
Those are wall-effect events which correspond to the flat region with lower
energies starting from 0.18 MeV in the spectrum.
The argon gas filled in the proportional counters has little effect on the neutron
detection efficiency but can significantly decrease the track lengths of
the produced proton and triton, and therefore reduce the wall effect. 
By defining a region of interest around the full energy peak between 0.64 - 0.8 MeV (ROI) as indicated in Fig.
\ref{fig_energy_spec} (a), we find 81.0$\pm$0.4\% of the detected events are located in it.

\section{Background measurements}
Stainless steel is chosen as the wall material of the proportional counters
because it has much lower $\alpha$ emission rate comparing to aluminum
\cite{bib:12,bib:13}.
The inset in Fig. \ref{fig_energy_spec} (a) shows an energy spectrum up
to 7 MeV measured at the CJPL.
The events located beyond the neutron full-energy peak are mainly due to
$\alpha$ radioactivity from the wall
material of the counters.
The $\alpha$ spectrum has a relatively flat distribution below $\sim$ 3 MeV and  shows a peak  at 5 MeV, a
typical energy of $\alpha$ decays from U/Th nuclides.
Below ~0.3 MeV energy region, lots of $\gamma$/$\beta$ events exist, overlapping
with the neutron events.
Following the procedure described in Ref. \cite{bib:12}, the $\beta/\gamma$
events can be well separated from neutrons in the raise time versus energy
two-dimensional plot as shown in Fig. \ref{fig_energy_spec} (b) taken at the
CJPL.

To characterize the background of the detector array,
ground- and underground-measurements were performed at the China
Institute of Atomic Energy and the A1 hall of the CJPL, respectively.
The neutron events were selected by using the ``neutron \& alpha" cut from the raise time versus energy
two-dimensional plot as shown in Fig. \ref{fig_energy_spec} (b).
Such events also include contributions from $\alpha$ background.
To disentangle the $\alpha$ backgrounds from the ``neutron \& alpha" events, a flat $\alpha$
energy spectrum is assumed \cite{bib:12,bib:13} in the energy range 0.18 - 2.0 MeV. 
The total background inside the "neutron \& alpha" cut as shown in Fig. \ref{fig_energy_spec} (b) measured at CJPL for the whole array is 4.5(2)/hour,
of which 1.94(5)/hour is from $\alpha$ background.
The above $\alpha$ background is evaluated using the $\alpha$ counts in the energy range 1.0 - 2.0 MeV which was assumed to have the same $\alpha$ emission rate as 0.18 - 0.8 MeV. This assumption was verified using a $^4$He proportional counter which will be discussed below. The total background is more than two orders of magnitude lower than the
result of 1238(11)/hour obtained from ground-measurement.

A $^{4}$He counter, which has exactly the same parameters as the
$^{3}$He counters but filled with $^{4}$He gas instead of $^{3}$He,
manufactured by the same supplier was used to investigate the $\alpha$
background in more detail.
The $^{4}$He counter is insensitive to neutrons but expected to have
similar $\alpha$ background shape as the $^{3}$He counters.
The $\alpha$ background rate, with 28.9-hour measurement time in CJPL,  was found to be 4.24 $\pm$ 0.48 and 4.22 $\pm$ 0.38 MeV$^{-1}$hour$^{-1}$, in 0.18 - 0.8 and 1.0 - 2.0 MeV energy regions, respectively. This justifies that the two energy regions have indeed the same $\alpha$ emission rates.
However, we found an $\alpha$ emission rate of 2.63(30)/hour by using the
same neutron cut as the $^{3}$He counters
in the raise time versus energy plot.
This is about a factor of 30 higher than the
$^{3}$He counters. Note that the $\alpha$ emission rate of 1.94(5)/hour quoted above includes the background of the 24 $^{3}$He counters while the 2.63(30)/hour is only for one $^{4}$He counter.
This implies that the $\alpha$ background could differ significantly even
for the same type of wall material. This has to be carefully investigated
at the manufacturing stage in order to achieve low $\alpha$ background.

To further reduce the $\alpha$ background, one can choose the events within the ROI only [see Fig. \ref{fig_energy_spec} (a)].
As a result, the $\alpha$ background is reduced by 74\% at the cost of
losing only about 19\% of the neutron detection efficiency.
This corresponds to a background of 2.5(1)/hour, of which 0.50(1)/hour is
from $\alpha$ background.
A better approach for $\alpha$-background suppression is to use plastic scintillator as the moderator, where
most of the $\alpha$ background can be eliminated via coincident
measurements \cite{bib:14}.
The $\alpha$ background is intrinsic properties of the proportional
counters used in the array which does not change with environments.
However, the neutron background may change with different environments and
therefore should be characterized for each use.

\section{Efficiency calibration}
\subsection{$^{51}$V(p, n)$^{51}$Cr Experiment and simulation}
\begin{figure}[hbt]
\includegraphics[width=0.4\textwidth]{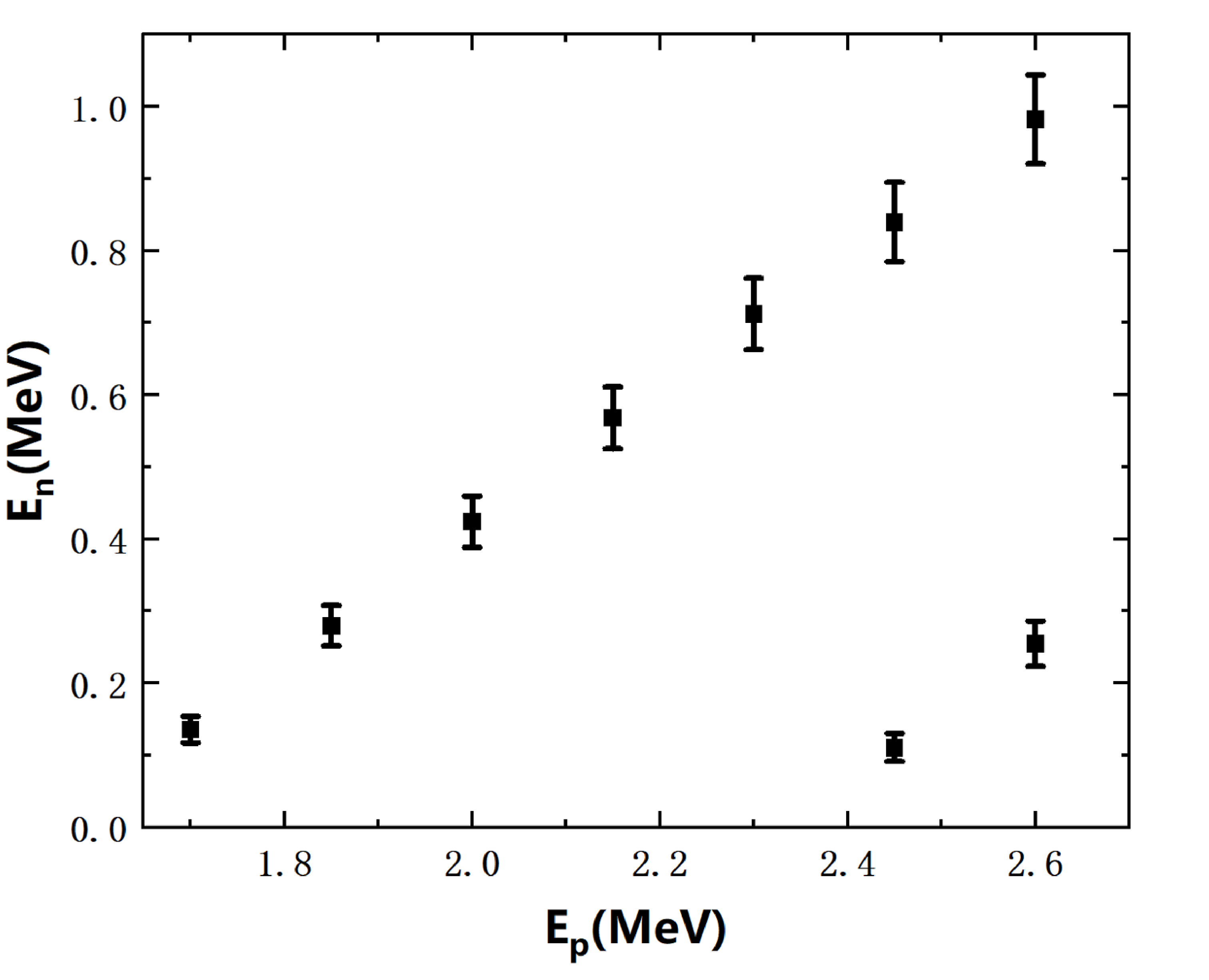}
	\caption{
		The emitted neutron energies from the $^{51}$V(p,n)$^{51}$Cr
		reaction as a function of proton beam energies. The height of the
		bars represents the energy spread of the emitted neutrons. The two components of
		the neutron energies at
		$E_p$ =  2.45 and 2.6 MeV  correspond to transitions feeding the
		ground and the first-excited states of $^{51}$Cr.
\label{fig_Ep_En}}
\end{figure}

Detection efficiency of the
array was calibrated at the nuclear physics experiment (NPE)
terminal of the 3-MV tandetron accelerator~\cite{bib:15} at Sichuan
university. The detection efficiency hereafter includes both full-energy and
wall-effect events. Quasi Mono-energetic neutrons were produced using the $^{51}$V(p,
n)$^{51}$Cr reaction (Q = -1534.8 keV) at incident energies between
1.7 and 2.6 MeV with a step of 0.15 MeV. The $^{51}$V(p, n)$^{51}$Cr reaction is widely used in
the calibration of neutron detectors~\cite{bib:11,bib:13,bib:16,bib:18,bib:19,bib:20} for its slow variation of neutron intensity and energy with
angle, and the well-known target preparation and utilization. Fig.
\ref{fig_Ep_En} shows the emitted neutron energy as a function of incident
proton beam energy. When the
incident energy is above 2.33 MeV, neutrons from transitions feeding the first excited state
of $^{51}$Cr ($E_{x} = $ 749 keV) are mixed with those feeding the ground
state. However, as pointed out in Ref.~\cite{bib:13}, the
contribution from transitions to the first excited state is negligible for
incident energies up to 2.6 MeV~\cite{bib:16}, which was also
confirmed in the present work (see discussions in the following sections).
The proton beams were focused at the target with a diameter of less than 5 mm
(full width at half maximum). The beam spots were monitored by a fluorescent
target together with a beam position monitor located 1.5 m upstream the
target for each beam energy.

Vanadium targets with a thickness of 110 $\mu g/cm^{2}$ were used in the
measurements. They were produced by evaporating  natural vanadium on 1-mm
thick tantalum disks with a diameter of 30 mm. The target thickness
corresponds to an energy loss of 8$\sim$11 keV for proton beams with
energies of 1.7 - 2.6 MeV. The beam intensity varied
from 4 $\mu$A to 120 nA as the beam energy increased from 1.7 to 2.6 MeV,
making
 the counting rate of the detector array stay around $10^4$/s with a dead time being less than
1\% for all beam energies. A direct water cooling of the
reaction target was used to
reduce the sputtering and target loss. The possible contribution of beam
induced background was investigated with a blank tantalum target at the beam
energy of 2.0 MeV.
The background contribution was found to be less than 1\%.

The total number of emitted neutrons was determined based on the activation
method as described in Ref.~\cite{bib:13}.
The number of radioactive product $^{51}$Cr equals to the number of emitted
neutrons in the reaction. $^{51}$Cr decays by electron capture with a half-life
of $T_{1/2}$ = 27.7025(24) days and has a branching ratio of $B$ = 9.91(1)\% to
decay to the first excited state of its daughter nuclei $^{51}$V, which is
followed by the emission of a 320-keV $\gamma$ ray. The off-line measurement of this
$\gamma$ ray was carried out with a GEM series HPGe detector whose
relative efficiency is 30\%. The absolute efficiency of the HPGe detector was measured at a
distance of 20 cm using $^{137}$Cs [1.534(19)$\times$10$^{5}$ Bq] and $^{152}
$Eu [5.72(6)$\times10^{4}$ Bq] $\gamma$-ray sources. The distance of 20 cm
is large enough to avoid pileups of cascading $\gamma$ rays from the
sources. A $^{51}$Cr $\gamma$ source was produced by the $^{51}$V(p, n)$^{51}$Cr
reaction. Its activity was measured at the position of 20 cm and then used to determine the efficiency for the 320-keV $\gamma$ line at the position of 10.2 cm ($\eta_{320}$),    where all the irradiated
targets were placed for off-line measurements.
$\eta_{320}$ was determined to be 0.498(7)\% at 10.2 cm.
As beam current was stable within the irradiation time, the number of emitted neutrons was then
determined by the off-line measurement of the 320 keV $\gamma$ ray with the
activation formula~\cite{bib:13}
\begin{equation}
	N_{R} = \frac{N_{\gamma}}{B\cdot\eta_{320}}\cdot\frac{e^{\lambda t_w}}{1-e^{-\lambda t_c}}\cdot\frac{\lambda\cdot t_i}{1-e^{-\lambda t_i}},
	\label{eq:eq_act_formula}
\end{equation}
where $N_{\gamma}$, $t_i$, $t_c$ and $t_w$ are the number of detected
320-keV $\gamma$ rays, the activation time, the counting time, and the
waiting time elapsed between the end of irradiation and the start of the
counting. $\lambda$ = ln(2)/$T_{1/2}$ is the decay constant of $^{51}$Cr.
After dead time correction, the detection efficiency
of the neutron detector is calculated as
\begin{equation}
	\eta_{n} = \frac{N_{n}}{N_{R}},
	\label{eq:eq_efficiency}
\end{equation}
where $N_n$ is the detected number of neutrons by the detector array.

Since the energy of the neutrons emitted in the $^{51}$V(p, n)$^{51}$Cr
reaction is only up to $\sim$1 MeV, it
is still far below that from $^{13}$C($\alpha$, n)$^{16}$O reaction. The Monte Carlo simulation code Geant4~\cite{bib:21,bib:22}
 of version 10.4.6 was used to determine the neutron detection efficiency at
neutron energies above 1 MeV.

\begin{figure}[hbt]
	\centering
	\includegraphics[width=0.4\textwidth]{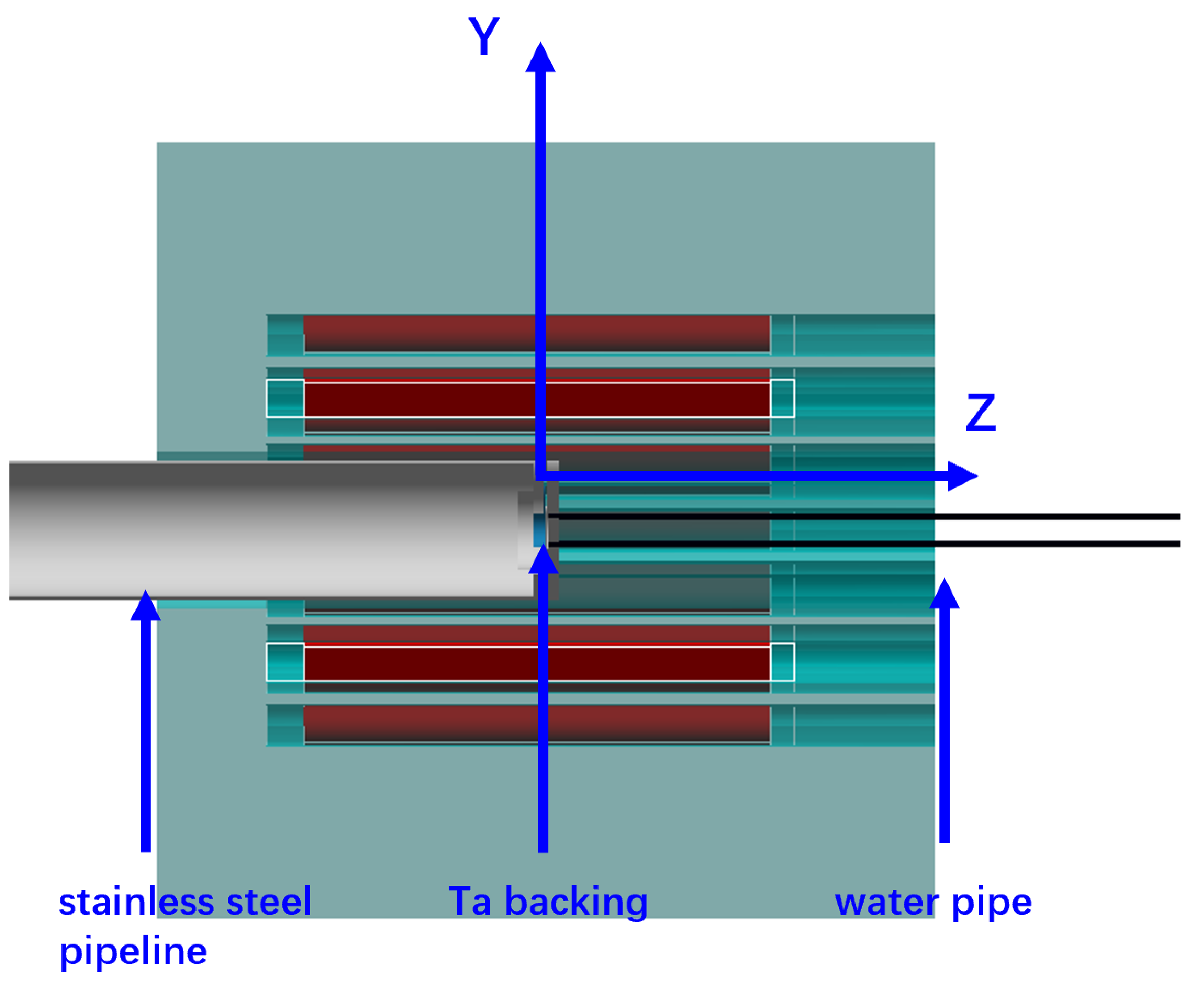}
	\caption{(Color online) The detector setup used in the simulations for
	the $^{51}$V(p, n)$^{51}$Cr reaction.
	  }
	\label{fig_g4_setup}
\end{figure}

\begin{figure*}[hbt]
	\centering
    \includegraphics[width=0.8\textwidth]{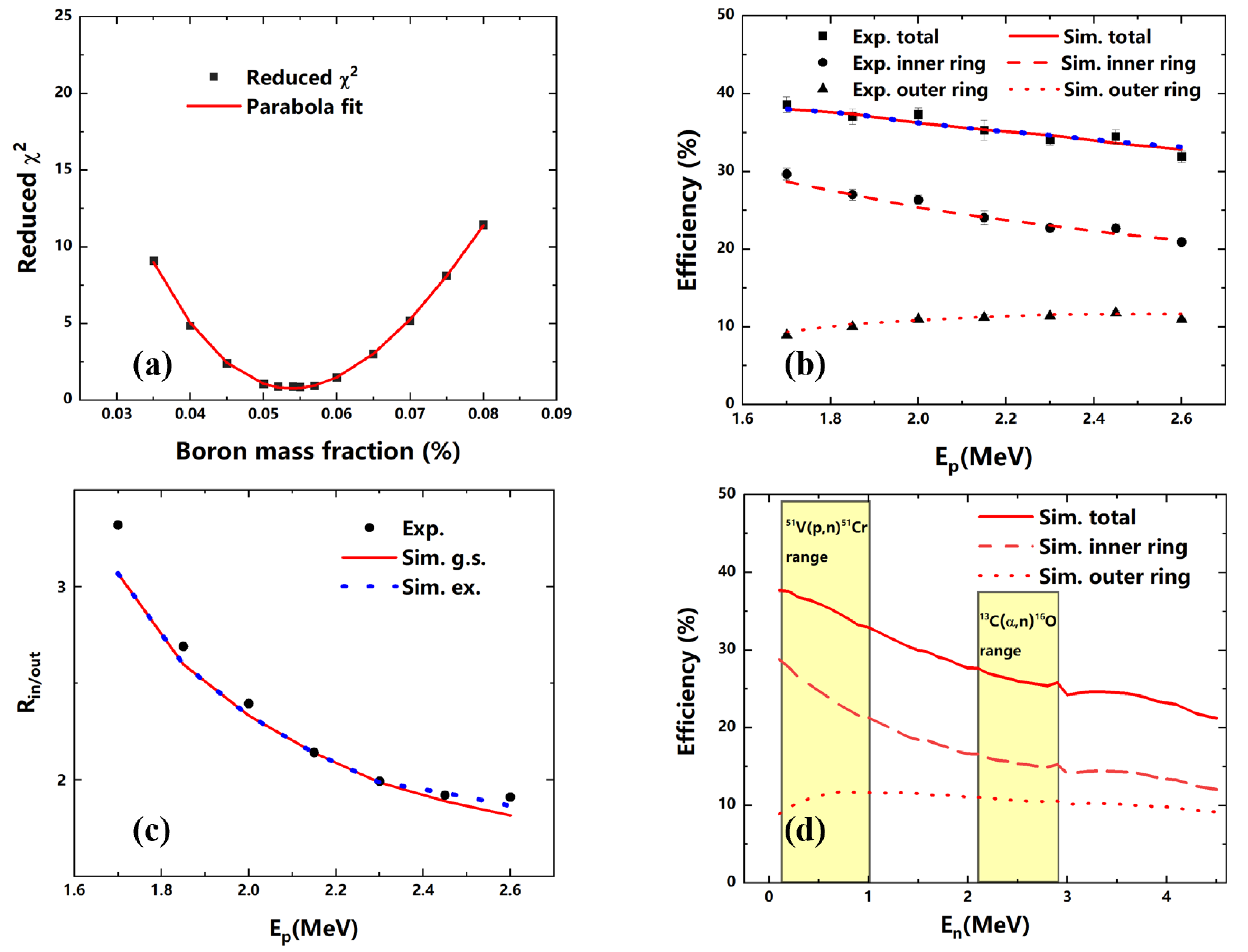}
	\caption{(Color online) (a) The reduced $\chi^2$, obtained by comparing the simulated and
	measured total detection efficiencies, as a function of boron mass fraction.
	The curve is a parabola fitting of the reduced $\chi^2$. (b) Measured and simulated total, inner ring and outer ring detection efficiencies as a function of proton beam energy in the $^{51}$V(p, n)$^{51}$Cr reaction. Blue short dashed curve represents the simulation with first excited state neutrons. The first excited state neutron component were estimated from a statistical model calculation, calibrated using the experimental data~\cite{bib:16}, using TALYS~\cite{bib:add4}. (c) Measured and simulated inner-to-outer-ring ratios $R_{in/out}$ at different proton beam energies. (d) Simulated total, inner ring and outer ring detection efficiencies for mono-energetic neutrons with an isotropic angular distribution. The neutron range indicated with $^{13}$C($\alpha$, n)$^{16}$O is for the $\alpha$ energy range $E_{\alpha}$ = 300 - 800 keV.
	  }
	\label{fig_chi2_eff_all}
\end{figure*}

The detailed detector setup used the simulation is shown in Fig.
\ref{fig_g4_setup}. The beam pipe,
target backing and water cooling loops were included in the simulation,
which resembles the real physical setups.
 An overestimation of
the neutron detection efficiency was obtained from the simulation
compared to the experimental results. Similar overestimations
have also been observed in several other setups~\cite{bib:11,bib:13}.
This overestimation is interpreted as that some of the
neutrons are absorbed by small contaminants in the moderating
polyethylene. Instead of using a normalization factor, we added a small amount of
boron into the moderating polyethylene in the simulation to take
into account the
neutron absorption effects.
Fig. \ref{fig_chi2_eff_all} (a) shows the reduced $\chi^2$, obtained by comparing the simulated and measured
detection efficiencies, as a function of boron mass fraction. The
minimum of $\chi^2$ was found at the boron mass fraction of 0.054\%
by fitting the curve using a parabola.
The simulated total, inner ring and outer ring detection efficiencies are
shown in Fig. \ref{fig_chi2_eff_all} (b). An estimation of the first excited state contributions was also carried out using the first excited state neutron component estimated from a statistic model calculation, calibrated using the experimental data~\cite{bib:16}, using TALYS~\cite{bib:add4}. The simulated total efficiencies with the first state neutron contributions are also shown as the blue short dashed curve in Fig.~\ref{fig_chi2_eff_all} (b). The maximum deviation between the simulation with and without the contribution of the first excited state is 0.9\%, which is ineligible by comparing to the experimental uncertainty of 3.7\%.

The boron contaminant affects
not only the total efficiency, but also the ratio $R_{in/out}$ of the detection
efficiencies of the inner and
the outer ring detectors.
With the boron mass fraction obtained above, the measured ratio $R_{in/out}$ is also
well reproduced by the two simulations (see Fig.~\ref{fig_chi2_eff_all} (c)). Except for the first point, the simulation results are 1\%-3\% lower than that of the experiment at $E_p<$ 2.4 MeV. Note that the isotropic angular distributions are used in the simulations. The discrepancy of $R_{in/out}$ between simulations and experiment is probably due to the non-isotropic distribution of the emitted neutrons. For $E_p$ = 2.45 and 2.6 MeV, the discrepancy between the simulated (with ground state neutrons)
and measured $R_{in/out}$ ratios starts to increase to 1.6\% and 5\%, respectively.
This is due to the opening of the
decay channel to the first excited state in $^{51}$Cr (see Fig.
\ref{fig_Ep_En}). Taking into account the first excited state neutrons estimated from TALYS, the discrepancy is reduced to 0.6\% and 2.3\% for $E_p$ = 2.45 and 2.6 MeV, respectively, as shown in Fig.~\ref{fig_chi2_eff_all} (c).

The contributions from the decay channel to the first excited
state is negligibly small regarding the total detection efficiency though
its effects on the $R_{in/out}$ is larger.
The total, inner ring and outer ring detection efficiencies
for mono-energetic neutrons up to 4.5 MeV with an isotropic angular distribution were also simulated and shown in Fig. ~\ref{fig_chi2_eff_all} (d).

\begin{figure}[hbt]
	\includegraphics[width=0.4\textwidth]{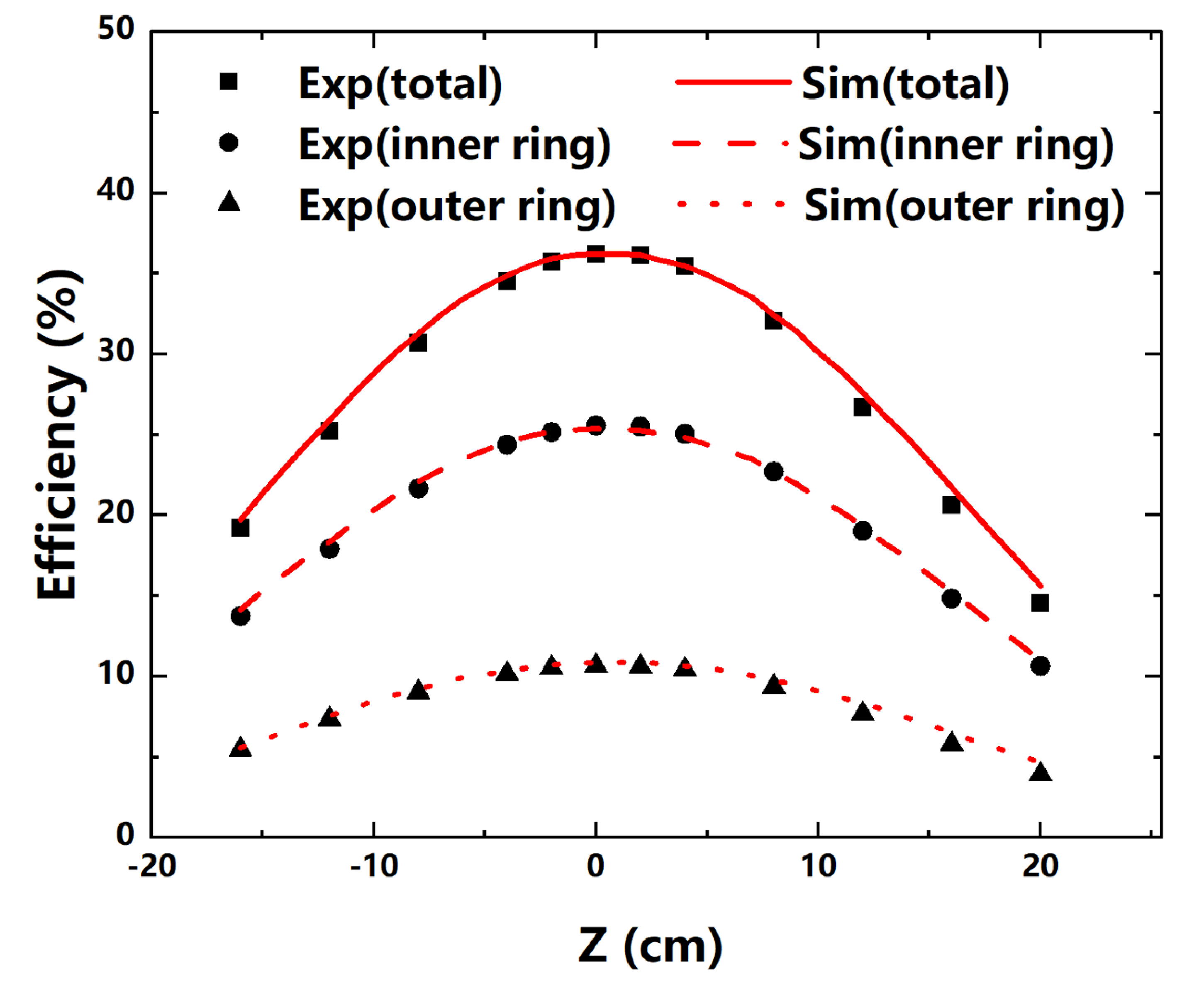}
	\caption{(Color online) Detection efficiencies of the total (solid  squares), the inner
	ring (solid  circles) and the outer ring (solid triangles) detectors as a
	function of the target position. The statistical errors are smaller than
	the symbol size. Solid, dashed and long dashed curves are those from the
	Geant4 simulations.}
	\label{fig:fig_Position}
\end{figure}

The dependence of detection efficiency on the source position was measured by placing the detector array
at different positions along the beam line ($Z$ axis)
using the $^{51}$V(p, n)$^{51}$Cr reaction at
$E_p$ = 2 MeV.
The position $Z$ = 0 corresponds to the target being at the center of the
array. Moving the detector array forward in the beam direction corresponds
to positive $Z$ values and backward  corresponds to  negative $Z$ values.
The relative efficiencies at different positions were normalized using integrated incident beam currents
on the target.
The results for the total, inner ring and outer ring detectors are shown in Fig.~\ref{fig:fig_Position} together with  the  Geant4 simulations.
In Fig. \ref{fig:fig_Position}, the
measured total efficiency at $Z$ = 0 was normalized to the simulated value.
Overall good agreements were found between the measurements and
 the  simulations.
 
Taking into account the average deviation between the experimental data and simulations between Fig.~\ref{fig_chi2_eff_all} (b) and Fig.~\ref{fig:fig_Position}, the difference between the Geant4 simulations and the experiment results is 2.8\%, which reflects one of the systematic uncertainties of our simulation and is quoted as the uncertainty of the Geant4 simulations.

\subsection{Extrapolating the detection efficiency for the study of the $^{13}$C($\alpha$, n)$^{16}$O reaction at stellar energies}
\begin{figure}[hbt]
	\centering
	\includegraphics[width=0.4\textwidth]{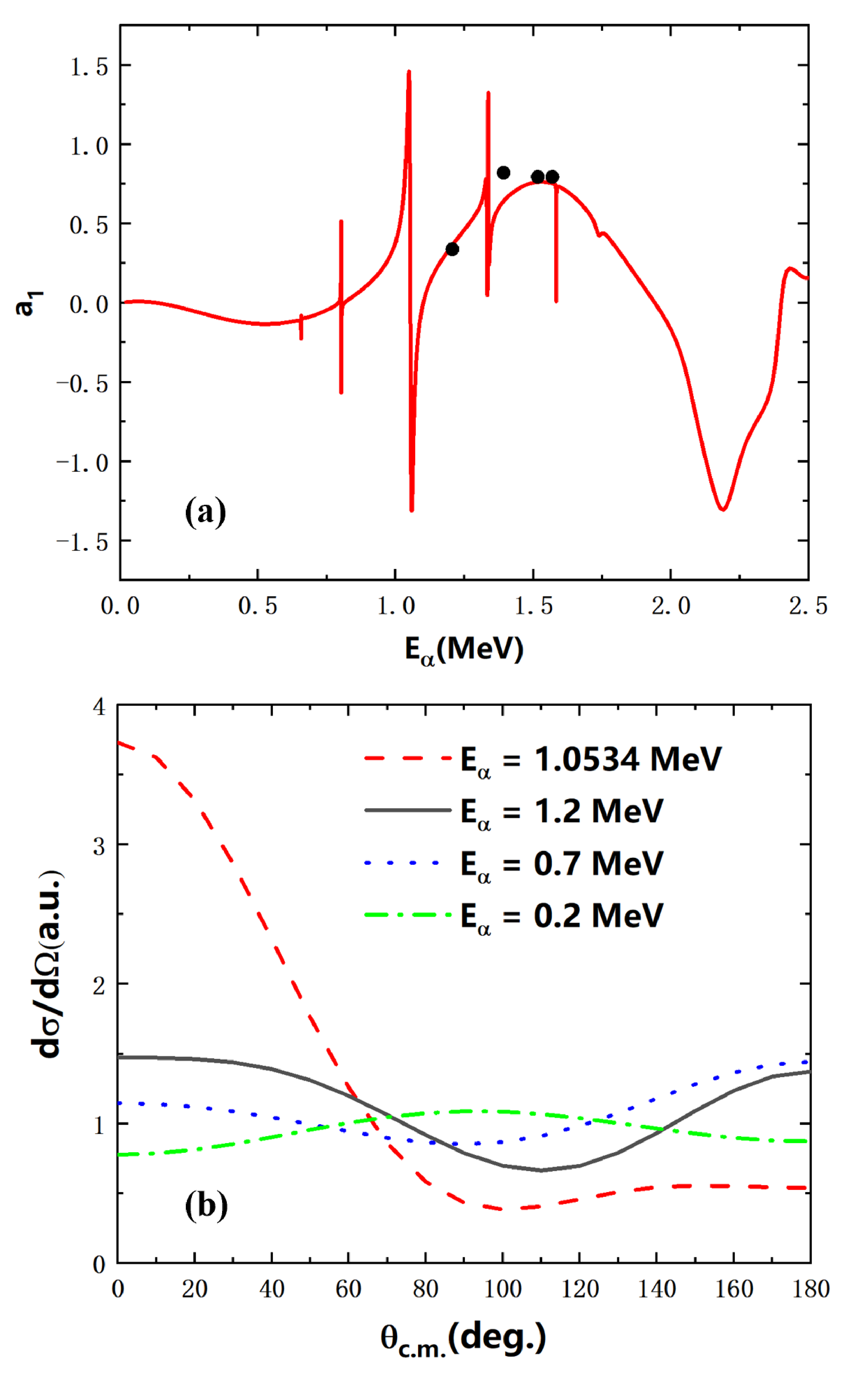}
	\caption{(Color online) (a) Legendre polynomials coefficient $a_1$ used in the
		angular distributions of the $^{13}$C($\alpha$, n)$^{16}$O reaction as a
		function of beam energies. The
		red curve is from Paris/Hale ENDF-8 \cite{bib:23} and the black dots are from Ref
		\cite{bib:24}. In order to match the excitation energy of $^{17}$O recommended by NNDC~\cite{bib:add5}, the Legendre polynomials coefficients around 1.05 MeV and 1.33 MeV are shifted 2.4 keV and 1.7 keV, respectively, to higher energy. (b) Representative angular distributions
		obtained at beam energies of 0.2, 0.7, 1.0534 and 1.2 MeV.}
	\label{fig:fig_Legendre}
\end{figure}

It should be noted that the simulated efficiency in Fig. \ref{fig_chi2_eff_all}
cannot be directly applied to the $^{13}$C($\alpha$, n)$^{16}$O reaction in
which the emitted neutrons are neither mono-energetic
nor isotropic. The asymmetry in the efficiency curve shown in Fig. \ref{fig:fig_Position} indicates
that the simulation predicts a slightly
higher detection efficiency for neutrons emitted at backward angles than
 that  at forward angles. 
Since the angular distribution of the $^{13}$C($\alpha$, n)$^{16}$O reaction
is not measured at low energies close to the Gamow window, theoretically
predicted angular distributions were used in our simulations.
Legendre polynomials up to the third order were used in calculating the
angular distributions. Fig.~\ref{fig:fig_Legendre} shows the Legendre polynomials coefficient $a_1$ used in the
angular distributions of the $^{13}$C($\alpha$, n)$^{16}$O reaction as a
function of beam energies in (a) and the representative angular distributions
obtained at beam energies of 0.2, 0.7, 1.0534 and 1.2 MeV in (b).
Adopting the angular
distributions in Refs. \cite{bib:23,bib:24,bib:25}, the angular distribution effect on the efficiencies is corrected in our Geant4 simulations.

\begin{figure}[hbt]
	\centering
    \includegraphics[width=0.4\textwidth]{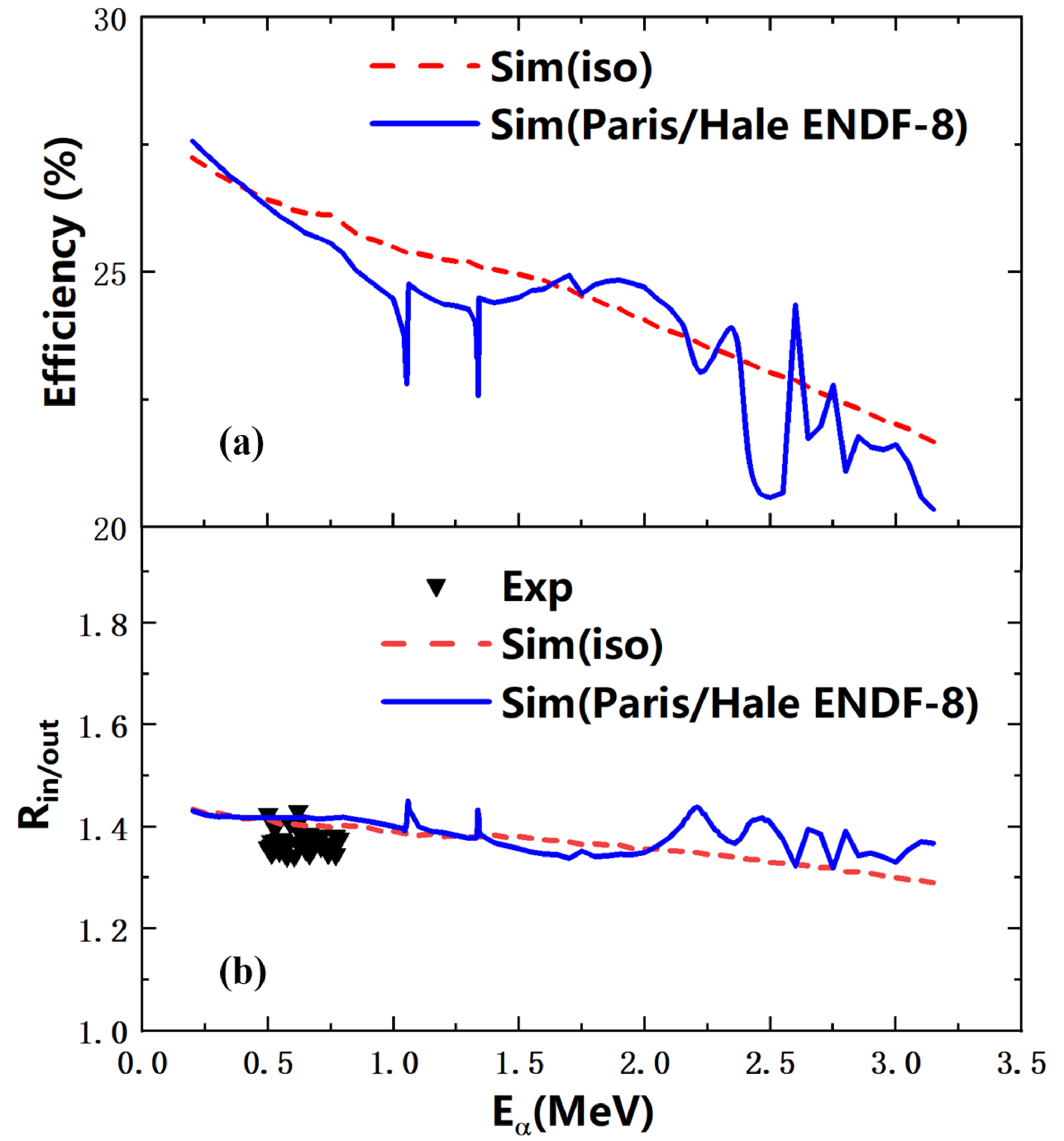}
	\caption{(Color online) (a) Simulated detection efficiencies using isotropic (red dashed line) and theoretically predicted (blue solid line) angular distributions. (b) Simulated ratios $R_{in/out}$ using both isotropic (red dashed line) and theoretically predicted (blue solid line) angular distributions. Solid triangles are the measured $R_{in/out}$ from Ref. \cite{bib:26}.
    \label{13C_a_n_eff}}
\end{figure}

Detection efficiencies were simulated using both isotropic
 and calculated angular distributions in the incident
$\alpha$ energies between 0.2 - 3.15 MeV.
 In the simulation, the narrow resonances are ignored. 
 Fig. \ref{13C_a_n_eff} (a) shows simulated
efficiencies using isotropic and calculated angular distributions.
For $\alpha$ energies between 0.3 - 0.8 MeV, the difference is 0.19\% - 2.28\% which
is relatively small.
However, above 0.8 MeV, the difference becomes larger with a maximum of 4\%. To evaluate the effect for the sharp resonance, we also simulated the detection efficiency at the energy where the $a_1$ coefficient reaches its maximum around E = 1.0534 MeV, as shown in Fig.~\ref{fig:fig_Legendre} (a). The resulting efficiency is 11\% less than the efficiency with an isotropic angular distribution. Therefore, the angular distribution effect needs to be evaluated carefully for the experiments aiming to a high precision.  

The effects of the angular distributions were further investigated by
comparing the simulated and measured ratios of the inner ring and outer ring detection
efficiencies $R_{in/out}$. As
 shown in Fig. \ref{13C_a_n_eff} (b), the $R_{in/out}$ is insensitive to the angular
 distribution and the simulations in both case agree reasonably well with the measured
 values from Ref. \cite{bib:26}. The difference between measured and simulated $R_{in/out}$ values is $\sim$ 4\%. An alternative way of extrapolating the total detection efficiency was carried out as follows. The inner-ring detection efficiency is extrapolated to the neutron energy range relevant to the $^{13}$C($\alpha$, n)$^{16}$O reaction. Then the total detection efficiency is obtained using the measured $R_{in/out}$ values from Ref. \cite{bib:26}. The difference in the total detection efficiency between the two extrapolation methods is $\sim$ 2.5\%, which is counted as another systematic uncertainty.


Taking into account the uncertainties of the Geant4 simulations (2.8\%), the extrapolation (2.5\%), the neutron angular distributions (2.3\%) and the detection efficiency of HPGe detector (1.5\%), the overall uncertainty of the detection efficiency for the $^{13}$C($\alpha$, n)$^{16}$O reaction in the energy range $E_\alpha$ = 300 - 800 keV is
determined to be 5\%. Excluding the narrow resonances, the maximum uncertainty of neutron angular distributions is 4\% if the energy range is extended up to 2.4 MeV. According to the ENDF angular distribution, the maximum uncertainty of neutron angular distributions at the energy beyond 2.4 MeV is 12\%.

\section{Summary}
A high-efficiency and low-background neutron detector array consisting of
24 $^{3}$He proportional counters embedded in a polyethylene moderator has been
developed for the cross-section measurement of the $^{13}$C($\alpha$,
n)$^{16}$O reaction at the China JinPing underground Laboratory. As a result of the deep underground location and a 5-cm thick borated polyethylene
shield, a low background of 4.5(2)/hour was achieved, of which 1.94(5)/hour
was from the internal $\alpha$ radioactivity.
The $^{51}$V(p, n)$^{51}$Cr reaction was used to calibrate the neutron
detection efficiency of the array for neutrons with energies $E_n$ $<$ 1 MeV.  For
$E_n$ $>$ 1 MeV, the Monte Carlo simulation code Geant4 was used to
extrapolate the detection efficiency. Specifically, an energy dependent
detection efficiency, which can be directly applied to the $^{13}$C($\alpha$,
n)$^{16}$O reaction, was obtained from the simulation.  The effects of the
angular distribution of the $^{13}$C($\alpha$,n)$^{16}$O reaction on the
detection efficiency, which were overlooked in previous works, were investigated in the present work and shown to be non-negligible.

\section*{Acknowledgments}
The authors thank the staff of the 3MV tandetron accelerator facility of Sichuan University for their support during the experiment. The authors also thank the assistance of Carl Brune in estimating the corrections of the neutron angular distribution. This work was supported by the National Natural Science Foundation of China under Grants No. 11490564 and 11805138.

\end{document}